\RequirePackage{fix-cm}
\documentclass[smallextended]{svjour3}       
\smartqed  
\usepackage{graphicx}
\usepackage{amssymb}
\begin{document}

\title{Multipartite unextendible entangled basis
}

\titlerunning{Multipartite unextendible basis}        

\author{Yu Guo         \and
       Yanping Jia   \and
        Xiulan Li
}

\authorrunning{Y. Guo et al.} 

\institute{Y. Guo, Y. Jia and X. Li \at
              School of Mathematics and Computer Science, Shanxi Datong University, Datong 037009, People's Republic of China \\
              \email{guoyu3@aliyun.com (Y. Guo)} }

\date{Received: date / Accepted: date}

\maketitle

\begin{abstract}
The unextendible entangled basis with any arbitrarily given Schmidt number $k$ (UEBk) in $\mathbb{C}^{d_1}\otimes\mathbb{C}^{d_2}$ is proposed in
[Phys. Rev. A 90 (2014) 054303], $1<k\leq \min\{d_1,d_2\}$, which is a set of orthonormal entangled states with Schmidt number $k$ in a $d_1\otimes d_2$ system consisting of fewer than
$d_1d_2$ vectors which have no additional entangled vectors with Schmidt number $k$ in the complementary space.
In this paper, we extend it to multipartite case and
a general way of constructing $(m+1)$-partite UEBk from $m$-partite UEBk is proposed ($m\geq 2$).
Consequently, we show that there are infinitely many UEBks in
$\mathbb{C}^{d_1}\otimes\mathbb{C}^{d_2}\otimes\cdots\otimes\mathbb{C}^{d_N}$ with any dimensions and any $N\geq3$.

\end{abstract}

\section{Introduction}

Entanglement and nonlocality are some of the most
indispensable concepts embodied in quantum physics
\cite{Einstein,Nielsen,Horodecki2009,Guhne}.
The nonlocal character of an entangled system
is the key to understanding the deepest implications of
quantum mechanics to information theory and even to the
nature of reality \cite{Bell}. It has been found that there are
sets of product states which nevertheless display a form of
nonlocality \cite{Bennett1999,DiVincenzo}.

One of the most interesting structures of the bipartite state space is that
there exists unextendible basis, which is impossible in a single Hilbert space corresponding to
a single particle.
The first unextendible basis proposed is the so-called unextendible product
basis (UPB) introduced firstly in Ref. \cite{Bennett1999}. UPB is a set of
incomplete orthogonal product pure states whose complementary space
does not contain product states.
It is shown that the members of a UPB are not perfectly distinguishable by local positive
operator valued measurements and classical communication, which shows the nonlocality without entanglement. Moreover, UPB can be used for
constructing bound entangled states \cite{Pittenger}.

The second unextendible basis is the unextendible maximally entangled basis (UMEB)~\cite{Bravyi}.
A UMEB is a set of orthonormal maximally entangled states in a two-qudit system consisting of fewer than
$d^2$ vectors which have no additional maximally entangled vectors orthogonal to all of them.
It is shown that there exists UMEB in any $d_1\otimes d_2$ system whenever $d_1\neq d_2$ \cite{Chen,Limaosheng}.
Recently, UPB and UMEB are extended to a more general case~\cite{Guowu2014}, UEBk, i.e.,
a set of orthonormal entangled states with Schmidt number $k$ in a $d_1\otimes d_2$ system consisting of fewer than
$d_1d_2$ vectors which have no additional entangled vectors with Schmidt number $k$ in the complementary space (
$1<k\leq \min\{d_1,d_2\}$).
It is proved that there are UEBks in any bipartite system~\cite{Guowu2014}.
Very recently, entangled basis with some fixed Schmidt number $k$ (EBk) is proposed~\cite{Guowu2015}.
We find that there exists EBk in any bipartite system and multipartite system although only rare multipartite pure states admit
the Schmidt decomposition form.
The UPB in multipartite has been investigated in Ref.~\cite{Bennett1999,DiVincenzo}.
In this paper,
we investigate the multipartite unextendible entangled basis with any fixed Schmidt number $k$ ($m$-partite UEBk).
Namely, we extend the unextendible entangled basis to multipartite system.
Consequently, we find a unified way of getting multipartite UEBks from that of the lower space,
which guarantees that UEBks exist in any system.
This provide us new tools in investigating quantum protocols associated with both bipartite and multipartite quantum systems.

The rest of this paper is constructed as follows. In
Sect. 2, we introduce some related notations and terminologies and then propose the definition of
$m$-partite UEBk.
Section 3 shows that there are infinitely many bipartite UEBks but not SUEBks using the similar scenario as that of Ref.~\cite{Guowu2015}.
Section 4 introduces our main result: any $(m+1)$-partite UEBk can be induced from an $m$-partite UEBk
and there are infinitely many UEBks in any multipartite system. In Sect. 5,
we list several examples of three-partite UEBks. Both SUEBk and UEBk but not SUEBk are proposed.
We conclude in Sect. 6 at last.

\section{Notations and terminologies}

For the sake of clarity, we recall the related definitions firstly.
A state $|\psi\rangle\in\mathbb{C}^{d_1}\otimes\mathbb{C}^{d_2}$ ($d_1\leq d_2$) is called a maximally entangled pure state if
it can be written as $|\psi\rangle=\frac{1}{\sqrt{d_1}}\sum_{i=0}^{d_1-1}|i_1\rangle|i_2\rangle$ for some orthonormal basis
$\{|i_1\rangle\}$ of $\mathbb{C}^{d_1}$ and some orthonormal set $\{|i_2\rangle\}$ of $\mathbb{C}^{d_2}$.

\smallskip

\noindent{\bf Definition~\cite{Chen}}  {\it A set of states $\{|\phi_i\rangle\in\mathbb{C}^{d_1}\otimes\mathbb{C}^{d_2}: i=1,2,\dots,n,n<d_1d_2\}$ is called a
$n$-member UMEB if and only if

(i) $|\phi_i\rangle$, $i=1,2,\dots,n$, are maximally entangled;

(ii) $\langle\phi_i|\phi_j\rangle=\delta_{ij}$;

(iii) if $\langle\phi_i|\psi\rangle=0$ for all $i=1,2,\dots,n$, then $|\psi\rangle$ cannot be maximally entangled.}

\smallskip

The Schmidt number of a pure state $|\psi\rangle\in\mathbb{C}^{d_1}\otimes\mathbb{C}^{d_2}$ is defined
as the length of the Schmidt decomposition \cite{Schmidt}:
if
$|\psi\rangle=\sum_{j=0}^{k-1}\lambda_{j}|e_j^{(1)}\rangle|e_j^{(2)}\rangle$ is its
Schmidt decomposition, then the Schmidt number of the pure state $|\psi\rangle$, denoted by
$S_r(|\psi\rangle)$, is $k$, i.e., $S_r(|\psi\rangle)=k$. It is known that
$S_r(|\psi\rangle)={\rm rank}(\rho_1)={\rm rank}(\rho_2)$,
where $\rho_i$ denotes the reduced state of the $i$-th part.

\smallskip

\noindent{\bf  Definition~\cite{Guowu2014}}  {\it A set of states $\{|\phi_i\rangle\in\mathbb{C}^{d_1}\otimes\mathbb{C}^{d_2}: i=1,2,\dots,n,n<d_1d_2\}$ is called a
$n$-member unextendible entangled bases with Schmidt number $k$ (UEBk) if and only if

(i) $S_r(|\phi_i\rangle)=k$, $i=1,2,\dots,n$;

(ii) $\langle\phi_i|\phi_j\rangle=\delta_{ij}$;

(iii) if $\langle\phi_i|\psi\rangle=0$ for all $i=1,2,\dots,n$, then $S_r(|\psi\rangle)\neq k$.}

\smallskip
A UEBk is a special UEBk (SUEBk) if all the Schmidt coefficients are equal to $1/\sqrt{k}$.
It is clear that UEBk reduces to UPB when $k=1$ while SUEBk reduces to UMEB when $k=d_1$.
Next, we extend the UEBk to multipartite case.
We call $|\psi\rangle\in\mathbb{C}^{d_1}\otimes\mathbb{C}^{d_2}\otimes\cdots\otimes\mathbb{C}^{d_m}$ has a Schmidt decomposition form if it can be written as \cite{Guowu2015}
\begin{eqnarray}
|\psi\rangle=\sum_{j=0}^{k-1}\lambda_j|e_j^{(1)}\rangle|e_j^{(2)}\rangle\cdots|e_j^{(m)}\rangle,
\end{eqnarray}
 where $\{|e_j^{(l)}\rangle\}$ is an orthonormal
set of $\mathbb{C}^{d_l}$, $\sum_j\lambda_j^2=1$, $\lambda_j >0$, $l=1$, 2, $\dots$, $m$, $m\geq 3$.
For convenience, we denote the length of the decomposition by $\tilde{S}_r(|\psi\rangle)$, i.e., $\tilde{S}_r(|\psi\rangle)=k$,
and we
still call $\tilde{S}_r(|\psi\rangle)$ the Schmidt number and call the coefficients $\lambda_j$s Schmidt coefficients.
It is easy to see that $|\psi\rangle\in\mathbb{C}^{d_1}\otimes\mathbb{C}^{d_2}\otimes\cdots\otimes\mathbb{C}^{d_N}$
admits the Schmidt decomposition form if and only if all the reduced states have the same eigenvalues
and the eigenvectors of all the bipartite or multipartite reduced states are fully separable.

We remark here that
the Schmidt decomposition is not valid for multipartite case in general. Only
rare pure states in the multipartite case admit the generalized
Schmidt decomposition
$|\psi\rangle=\sum_{j=0}^{k-1}$
$
\lambda_j|e^{(1)}_j\rangle|e^{(2)}_j\rangle\otimes\cdots\otimes|e^{(m)}_j\rangle$
\cite{Peres1995,Thapliyal}, where $k\leq \min\{d_1,d_2,\dots,d_m\}$ and $d_i$ denotes the dimension of the
$i$-th subsystem. For the simplest three-partite case, any
three-qubit pure state admits the form  $|\psi\rangle
=\lambda_0|000\rangle+\lambda_1e^{i\theta}|100\rangle
+\lambda_2|101\rangle+\lambda_3|110\rangle +\lambda_4|111\rangle$\cite{Acin},
where $\lambda_i\geq0$, $\sum_i\lambda_i^2=1$,
$\theta\in[0,\pi]$, which is not a form of Schmidt
decomposition.

We now give the definition of $N$-partite unextendible EBk, which is a generalization of bipartite UEBk in Ref.~\cite{Guowu2014}.
We always assume with no loss of generality that $d_1\leq d_2\leq\cdots\leq d_N$ throughout this paper for simplicity.

\smallskip

\noindent{\bf  Definition 1}  {\it An orthonormal set $\{|\psi_i\rangle:i=1,2,\dots,n, n<d_1d_2\cdots d_m\}$ in $\mathbb{C}^{d_1}\otimes\mathbb{C}^{d_2}\otimes\cdots\otimes\mathbb{C}^{d_m}$ is
a $n$-member $m$-partite UEBk ($1\leq k\leq d_1$) if

(i) $\tilde{S}_r(|\phi_i\rangle)=k$, $i=1,2,\dots,n$;

(ii) $\langle\phi_i|\phi_j\rangle=\delta_{ij}$;

(iii) if $\langle\phi_i|\psi\rangle=0$ for all $i=1,2,\dots,n$, then either $|\psi\rangle$ does not admit a Schmidt decomposition form or $\tilde{S}_r(|\psi\rangle)\neq k$. Particularly, it is a $n$-member $m$-partite SUEBk if it is an $m$-partite UEBk with all the coefficients $\lambda_j^{(i)}$s of $|\psi_i\rangle$ equal to $\frac{1}{\sqrt{k}}$, namely, $|\psi_i\rangle=\sum_{j=0}^{k-1}\frac{1}{\sqrt{k}}|e_j^{(1)}\rangle|e_j^{(2)}\rangle\cdots|e_j^{(N)}\rangle$.}

\smallskip

Let $\mathcal{M}_{d_1\times d_2}$ be the space of all $d_1$ by $d_2$ complex matrices.
Then $\mathcal{M}_{d_1\times d_2}$ is a Hilbert space equipped with the inner product defined by $\langle A|B\rangle={\rm Tr}(A^{\dag}B)$ for any $A$, $B\in\mathcal{M}_{d_1\times d_2}$.
$\{A_i:{\rm rank}(A_i)=k, \ {\rm Tr}(A_i^{\dag}A_j)=\delta_{ij}, i=1,2,\dots,d_1d_2\}$ is called a rank-$k$ Hilbert-Schmidt basis of $\mathcal{M}_{d_1\times d_2}$ \cite{Guowu2015}.
There is a one-to-one relation between the EBk $\{|\psi_i\rangle\}$ and the rank-$k$ Hilbert-Schmidt basis $\{A_i\}$ \cite{Guowu2015}:
\begin{eqnarray}
|\psi_i\rangle=\sum_{k,l}a^{(i)}_{kl}|k_1\rangle|l_2\rangle\in\mathbb{C}^{d_1}\otimes\mathbb{C}^{d_2}\Leftrightarrow A_i=[a_{kl}^{(i)}]\in\mathcal{M}_{d_1\times d_2}, \nonumber\\
S_r(|\psi_i\rangle)={\rm rank}(A_i),\
\langle\psi_i|\psi_j\rangle={\rm Tr}(A_i^\dag A_j),~~~~~~~~~~~
\label{relation}
\end{eqnarray}
where $\{|i_1\rangle\}$ and
$|i_2\rangle$ are the standard computational bases of $\mathbb{C}^{d_1}$ and $\mathbb{C}^{d_2}$,
respectively.
For simplicity, we give the following definition.

\smallskip

\noindent{\bf  Definition 2}  {\it A set of $d_1\times d_2$ matrices  $\{A_i:i=1,2,\dots,n, n<d_1d_2\}$ is called an unextendible rank-$k$ Hilbert-Schmidt basis of $\mathcal{M}_{d_1\times d_2}$
if

i) ${\rm rank}(A_i)=k$ for any $i$;

ii) ${\rm Tr}(A_i^{\dag}A_j)=\delta_{ij}$;

iii) if ${\rm Tr}(A_i^{\dag}B)=0$, $i=1$, 2, $\dots$, $n$, then ${\rm rank}(B)\neq k$.}

\smallskip

It turns out that $\{A_i:{\rm rank}(A_i)=k\}$ is an unextendible Hilbert-Schmidt basis of $\mathcal{M}_{d_1\times d_2}$ if and only if $\{|\psi_i\rangle\}$
is a UEBk of $\mathbb{C}^{d_1}\otimes\mathbb{C}^{d_2}$.
Therefore, the UEBk problem is equivalent to the unextendible rank-$k$ Hilbert-Schmidt basis of the associated matrix space.

Any $m$-partite pure state $|\psi\rangle$ in $\mathbb{C}^{d_1}\otimes\mathbb{C}^{d_2}\otimes\cdots\otimes\mathbb{C}^{d_m}$ can be represented by
\begin{eqnarray}
|\psi\rangle=\sum\limits_{i_1,i_2,\dots,i_m}a_{i_1i_2\cdots i_m}|i_1\rangle|i_2\rangle\cdots|i_m\rangle,
\end{eqnarray}
where $\{|i_l\rangle\}$ and
is standard computational basis of $\mathbb{C}^{d_l}$, $l=1$, 2, $\dots$, $m$.
The space $\mathbb{C}^{d_1}\otimes\mathbb{C}^{d_2}\otimes\cdots\otimes\mathbb{C}^{d_m}$
can be regarded as a bipartite space for any bipartite cutting, thus the coefficients $a_{i_1i_2\cdots i_m}$
can be viewed as the corresponding matrix entries.
For example, we take $m=4$, the space is cut as $12|34$, then $a_{i_1i_2i_3i_4}$ is
$a_{i_1i_2|i_3i_4}$ indeed.
It follows that $[a_{i_1i_2|i_3i_4}]$ is a $d_1d_2\times d_3d_4$ matrix when we
regard both $i_1i_2$ and $i_3i_4$ as the single indexes respectively.
That is, any multipartite state corresponds to a matrix with respect to some bipartite cutting of the given space.
However, the unextendible rank-$k$ Hilbert-Schmidt basis of the matrix space corresponding to the multipartite state space can not induce
a UEBk in general.
For example,
\begin{eqnarray}
A_1=\frac{1}{3}\left(\begin{array}{cccc}
-1&0&0&0\\
2&2&0&0\end{array}\right),
A_2=\frac{1}{3}\left(\begin{array}{cccc}
2&0&0&0\\
-1&2&0&0\end{array}\right),\nonumber\\
A_3=\frac{1}{3}\left(\begin{array}{cccc}
2&0&0&0\\
2&-1&0&0\end{array}\right),
A_4=\frac{1}{3}\left(\begin{array}{cccc}
0&0&-1&0\\
0&0&2&2\end{array}\right),\nonumber\\
A_5=\frac{1}{3}\left(\begin{array}{cccc}
0&0&2&0\\
0&0&-1&2\end{array}\right),
A_6=\frac{1}{3}\left(\begin{array}{cccc}
0&0&2&0\\
0&0&2&-1\end{array}\right)~
\end{eqnarray}
form an unextendible rank-2 Hilbert-Schmidt basis of
$\mathcal{M}_{2\times(2\times2)}$ (corresponding to the bipartite cutting $1|23$).
However the corresponding set of pure state $\{|\psi_i\rangle:i=1,2,\dots,6\}$ is not a UEB2 of $\mathbb{C}^2\otimes \mathbb{C}^2\otimes \mathbb{C}^2$.

\section{UEBk but not SUEBk for bipartite case}

In Ref.~\cite{Guowu2014}, we showed that, in $\mathbb{C}^{d_1}\otimes \mathbb{C}^{d_2},$ there are at least $k-r$ (here $r=d_1$ mod $k$, or $r=d_2$ mod $k$) sets of UEBk when $d_1$ or $d_2$ is not the multiple of $k$, while
there are at least $2(k-1)$ sets of UEBk when both $d_1$ and $d_2$ are the multiples of $k$.
Note that the UEBks there are in fact SUEBks. In this section, we show that there are UEBks but not SUEBks in any bipartite system.
We always assume that $d_1\leq d_2$ unless otherwise specified.

For any possible $k$, if $d_1=sk+r$ and $d_2=s'k+r'$, $0\leq r<k$, $0\leq r'<k$, then
the Hilbert space $\mathcal{M}_{d_1\times d_2}$
is a direct sum of three subspaces which are equivalent to
$\mathcal{M}_{d_1\times (s'-1)k}$, $\mathcal{M}_{(s-1)k\times (k+r')}$ and $\mathcal{M}_{(k+r)\times (k+r')}$ respectively.
Since there always exist unextendible rank-$k$ Hilbert-Schmidt basis
in both $\mathcal{M}_{d_1\times (s'-1)k}$ and $\mathcal{M}_{(s-1)k\times (k+r')}$ \cite{Guowu2015},
we thus only need to check whether there exists unextendible rank-$k$ Hilbert-Schmidt basis in $\mathcal{M}_{(k+r)\times (k+r')}$.
We begin with the case of $k=2$.
There are three cases: $r=r'=0$; $r=0$ and $r'=1$ (or $r=1$ and $r'=0$); $r=r'=1$.
Therefore, we only need to consider
$\mathcal{M}_{2\times 2}$, $\mathcal{M}_{2\times 3}$ and $\mathcal{M}_{3\times 3}$.
Note that
\begin{eqnarray}
\left(\begin{array}{cc}
*&*\\
{*}&*\end{array}\right)=
\left(\begin{array}{cc}
*&0\\
{*}&*\end{array}\right)\oplus\left(\begin{array}{cc}
0&*\\
0&0\end{array}\right)
\end{eqnarray}
and the rank-2 Hilbert-Schmidt basis of
$\mathcal{L}_{3}^2=\left(\begin{array}{cc}
*&0\\
{*}&*\end{array}\right)$ is equivalent to that of
$\left(\begin{array}{cc}
*&0\\
0&*\\
0&*\end{array}\right)$ which is discussed in Ref.~\cite{Guowu2014}.
Using the method in Ref.~\cite{Guowu2014}, the following three vectors form a UEB2 in $\mathbb{C}^2\otimes\mathbb{C}^2$
\begin{eqnarray}
|\psi_0\rangle&=&\frac{1}{3}(2|1\rangle|0\rangle+2|1\rangle|1\rangle-|0\rangle|0\rangle),\nonumber\\
|\psi_1\rangle&=&\frac{1}{3}(2|0\rangle|0\rangle+2|1\rangle|1\rangle-|1\rangle|0\rangle),\nonumber\\
|\psi_2\rangle&=&\frac{1}{3}(2|0\rangle|0\rangle+2|1\rangle|0\rangle-|1\rangle|1\rangle),
\label{2qubits}
\end{eqnarray}
since the corresponding rank-2 Hilbert-Schmidt basis is
\begin{eqnarray}
A_1=\frac{1}{3}\left(\begin{array}{cc}
-1&0\\
2&2\end{array}\right),
A_2=\frac{1}{3}\left(\begin{array}{cc}
2&0\\
-1&2\end{array}\right),
A_3=\frac{1}{3}\left(\begin{array}{cc}
2&0\\
2&-1\end{array}\right),
\label{rank2}
\end{eqnarray}
which is unextendible.
In fact, any $3$ by $3$ isometric matrix  $X=[x_{kl}]$ without zero entries can induce a UEB2 of $\mathbb{C}^2\otimes\mathbb{C}^2$ since
we can replace the entries of $A_i$ in Eq.~(\ref{rank2}) by the entries in the $i$-th column of any $3$ by $3$ isometric matrix without zero entries, respectively.
(An $m$ by $n$ matrix $A$ is an isometric matrix if $A^\dag A=I_n$, $I_n$ is the $n$ by $n$ identity matrix.)
There are infinitely many UEB2s in $\mathbb{C}^2\otimes\mathbb{C}^2$ due to the fact that infinitely many isometric matrices exist.
The space of $2\times 3$ matrix space can be decomposed as
\begin{eqnarray*}
\left(\begin{array}{ccc}
*&*&*\\
{*}&*&*\end{array}\right)=
\left(\begin{array}{ccc}
{*}&*&0\\
{*}&*&0\end{array}\right)\oplus\left(\begin{array}{ccc}
{0}&0&*\\
{0}&0&*\end{array}\right).
\end{eqnarray*}
Since $\left(\begin{array}{ccc}
{*}&*&0\\
{*}&*&0\end{array}\right)$ have rank-2 Hilbert-Schmidt bases \cite{Guowu2015}
and any nonzero matrix in $\left(\begin{array}{ccc}
{0}&0&*\\
{0}&0&*\end{array}\right)$ is of rank-one and is orthogonal to that of
$\left(\begin{array}{ccc}
{*}&*&0\\
{*}&*&0\end{array}\right)$,
we conclude that there are (infinitely many) UEB2s in $\mathbb{C}^2\otimes\mathbb{C}^3$.
The following are three types of decomposition of $\mathcal{M}_{3\times 3}$
\begin{eqnarray*}
\begin{array}{lll}
\left(\begin{array}{ccc}
*&*&*\\
{*}&*&*\\
{*}&*&*\end{array}\right)&=&
\left(\begin{array}{ccc}
{*}&*&*\\
{*}&*&*\\
0&0&0\end{array}\right)
\oplus\left(\begin{array}{ccc}
0&0&0\\
0&0&0\\
{*}&*&*\end{array}\right),\\
\left(\begin{array}{ccc}
*&*&*\\
{*}&*&*\\
{*}&*&*\end{array}\right)&=&
\left(\begin{array}{ccc}
0&0&0\\
{*}&0&0\\
{*}&*&0\end{array}\right)
\oplus\left(\begin{array}{ccc}
*&0&0\\
0&*&0\\
0&0&0\end{array}\right)
\oplus\left(\begin{array}{ccc}
0&*&0\\
0&0&*\\
0&0&0\end{array}\right)
\oplus\left(\begin{array}{ccc}
0&0&*\\
0&0&0\\
0&0&*\end{array}\right),\\
\left(\begin{array}{ccc}
*&*&*\\
{*}&*&*\\
{*}&*&*\end{array}\right)&=&
\left(\begin{array}{ccc}
{*}&*&0\\
{*}&*&0\\
0&0&0\end{array}\right)
\oplus\left(\begin{array}{ccc}
0&0&0\\
0&0&*\\
0&*&0\end{array}\right)
\oplus\left(\begin{array}{ccc}
0&0&*\\
0&0&0\\
{*}&0&0\end{array}\right)
\oplus\left(\begin{array}{ccc}
0&0&0\\
0&0&0\\
0&0&*\end{array}\right).
\end{array}
\end{eqnarray*}
We denote the ten subspaces at the right hand of the three equations above by
$\mathcal{L}_{1,1}^{(2)}$, $\mathcal{L}_{1,2}^{(2)}$,
$\mathcal{L}_{2,1}^{(2)}$, $\mathcal{L}_{2,2}^{(2)}$, $\mathcal{L}_{2,3}^{(2)}$, $\mathcal{L}_{2,4}^{(2)}$,
$\mathcal{L}_{3,1}^{(2)}$, $\mathcal{L}_{3,2}^{(2)}$, $\mathcal{L}_{3,3}^{(2)}$ and $\mathcal{L}_{3,4}^{(2)}$, respectively.
Namely $\mathcal{L}_{i,j}^{(2)}$ denotes the $j$-th subspace of the $i$-th equation.
It is obvious that i) $\mathcal{L}_{i,k}^{(2)}$ is orthogonal to $\mathcal{L}_{i,l}^{(2)}$ whenever $k\neq l$,
ii)
$\mathcal{L}_{1,1}^{(2)}$,
$\mathcal{L}_{2,1}^{(2)}$, $\mathcal{L}_{2,2}^{(2)}$, $\mathcal{L}_{2,3}^{(2)}$,
$\mathcal{L}_{3,1}^{(2)}$, $\mathcal{L}_{3,2}^{(2)}$ and $\mathcal{L}_{3,3}^{(2)}$
have rank-2 Hilbert-Schmidt bases,
and iii) any nonzero matrix in $\mathcal{L}_{1,2}^{(2)}$,
or $\mathcal{L}_{2,4}^{(2)}$, or $\mathcal{L}_{3,4}^{(2)}$ is of rank-one.
One can check that each rank-2 space above
has infinitely many rank-2 Hilbert-Schmidt basis due to the infinitely many isometric matrices with
the corresponding size.
This implies that there are infinitely many 6-member, 7-member and 8-member UEB2s in $3\otimes 3$ system.

If $k=3$, we only need to discuss the spaces
$\mathcal{M}_{3\times 3}$, $\mathcal{M}_{3\times 4}$, $\mathcal{M}_{3\times 5}$,
$\mathcal{M}_{4\times 4}$, $\mathcal{M}_{4\times 5}$ and $\mathcal{M}_{5\times 5}$.
We decompose them into subspaces with dimension 3 and subspaces with dimension $l$, $l<3$, as following.
\begin{eqnarray*}
\left(\begin{array}{ccc}
*&*&*\\
{*}&*&*\\
{*}&*&*\end{array}\right)=
\left(\begin{array}{ccc}
0&0&*\\
{*}&0&0\\
{*}&*&0\end{array}\right)
\oplus\left(\begin{array}{ccc}
*&0&0\\
0&*&0\\
0&0&*\end{array}\right)\oplus\left(\begin{array}{ccc}
0&*&0\\
0&0&*\\
0&0&0\end{array}\right),
\end{eqnarray*}
\begin{eqnarray*}
\begin{array}{lll}
\left(\begin{array}{cccc}
*&*&*&*\\
{*}&*&*&*\\
{*}&*&*&*\end{array}\right)&=&
\left(\begin{array}{cccc}
0&0&0&*\\
{*}&0&0&0\\
{*}&*&0&0\end{array}\right)
\oplus\left(\begin{array}{cccc}
*&0&0&0\\
0&*&0&0\\
0&0&*&0\end{array}\right)
\oplus\left(\begin{array}{cccc}
0&*&0&0\\
0&0&*&0\\
0&0&0&*\end{array}\right)
\oplus\left(\begin{array}{cccc}
0&0&*&0\\
0&0&0&*\\
0&0&0&0\end{array}\right),
\end{array}
\end{eqnarray*}
\begin{eqnarray*}
\begin{array}{lll}
\left(\begin{array}{ccccc}
*&*&*&*&*\\
{*}&*&*&*&*\\
{*}&*&*&*&*\end{array}\right)&=&
\left(\begin{array}{ccccc}
0&0&0&0&*\\
{*}&0&0&0&0\\
{*}&*&0&0&0\end{array}\right)
\oplus\left(\begin{array}{ccccc}
*&0&0&0&0\\
0&*&0&0&0\\
0&0&*&0&0\end{array}\right)\oplus\left(\begin{array}{ccccc}
0&*&0&0&0\\
0&0&*&0&0\\
0&0&0&*&0\end{array}\right)\\
&&
\oplus\left(\begin{array}{ccccc}
0&0&*&0&0\\
0&0&0&*&0\\
0&0&0&0&*\end{array}\right)
\oplus\left(\begin{array}{ccccc}
0&0&0&*&0\\
0&0&0&0&*\\
0&0&0&0&0\end{array}\right),
\end{array}
\end{eqnarray*}
\begin{eqnarray*}
\begin{array}{lll}
\left(\begin{array}{cccc}
*&*&*&*\\
{*}&*&*&*\\
{*}&*&*&*\\
{*}&*&*&*\end{array}\right)&=&
\left(\begin{array}{ccccc}
0&0&0&*\\
0&0&0&0\\
{*}&0&0&0\\
{*}&*&0&0\end{array}\right)
\oplus\left(\begin{array}{cccc}
0&0&0&0\\
{*}&0&0&0\\
0&*&0&0\\
0&0&*&0\end{array}\right)
\oplus\left(\begin{array}{cccc}
*&0&0&0\\
0&*&0&0\\
0&0&*&0\\
0&0&0&0\end{array}\right)\\
&&
\oplus\left(\begin{array}{cccc}
0&*&0&0\\
0&0&*&0\\
0&0&0&*\\
0&0&0&0\end{array}\right)
\oplus\left(\begin{array}{cccc}
0&0&*&0\\
0&0&0&*\\
0&0&0&0\\
0&0&0&0\end{array}\right),
\end{array}
\end{eqnarray*}
\begin{eqnarray*}
\begin{array}{lll}
\left(\begin{array}{ccccc}
*&*&*&*&*\\
{*}&*&*&*&*\\
{*}&*&*&*&*\\
{*}&*&*&*&*\end{array}\right)&=&
\left(\begin{array}{ccccc}
0&0&0&0&*\\
0&0&0&0&0\\
{*}&0&0&0&0\\
{*}&*&0&0&0\end{array}\right)
\oplus\left(\begin{array}{ccccc}
0&0&0&0&0\\
{*}&0&0&0&0\\
0&*&0&0&0\\
0&0&*&0&0\end{array}\right)
\oplus\left(\begin{array}{ccccc}
*&0&0&0&0\\
0&*&0&0&0\\
0&0&*&0&0\\
0&0&0&0&0\end{array}\right)\\
&&
\oplus\left(\begin{array}{ccccc}
0&*&0&0&0\\
0&0&*&0&0\\
0&0&0&*&0\\
0&0&0&0&0\end{array}\right)
\oplus\left(\begin{array}{ccccc}
0&0&*&0&0\\
0&0&0&*&0\\
0&0&0&0&*\\
0&0&0&0&0\end{array}\right)
\oplus\left(\begin{array}{ccccc}
0&0&0&*&0\\
0&0&0&0&*\\
0&0&0&0&0\\
0&0&0&0&0\end{array}\right),
\end{array}
\end{eqnarray*}
\begin{eqnarray*}
\left(\begin{array}{ccccc}
*&*&*&*&*\\
{*}&*&*&*&*\\
{*}&*&*&*&*\\
{*}&*&*&*&*\\
{*}&*&*&*&*\end{array}\right)&=&
\left(\begin{array}{ccccc}
0&0&0&0&*\\
0&0&0&0&0\\
0&0&0&0&0\\
{*}&0&0&0&0\\
{*}&*&0&0&0\end{array}\right)
\oplus\left(\begin{array}{ccccc}
0&0&0&0&0\\
0&0&0&0&0\\
{*}&0&0&0&0\\
0&*&0&0&0\\
0&0&*&0&0\end{array}\right)
\oplus\left(\begin{array}{ccccc}
0&0&0&0&0\\
{*}&0&0&0&0\\
0&*&0&0&0\\
0&0&*&0&0\\
0&0&0&0&0\end{array}\right)
\oplus\left(\begin{array}{ccccc}
{*}&0&0&0&0\\
0&*&0&0&0\\
0&0&*&0&0\\
0&0&0&0&0\\
0&0&0&0&0\end{array}\right)\\
&&\oplus\left(\begin{array}{ccccc}
0&*&0&0&0\\
0&0&*&0&0\\
0&0&0&*&0\\
0&0&0&0&0\\
0&0&0&0&0\end{array}\right)
\oplus\left(\begin{array}{ccccc}
0&0&*&0&0\\
0&0&0&*&0\\
0&0&0&0&*\\
0&0&0&0&0\\
0&0&0&0&0\end{array}\right)
\oplus\left(\begin{array}{ccccc}
0&0&0&*&0\\
0&0&0&0&*\\
0&0&0&0&0\\
0&0&0&0&0\\
0&0&0&0&0\end{array}\right).
\end{eqnarray*}
For more simplicity, we denote the 30 subspaces at the right hand of the three equations above by
$\mathcal{L}_{1,1}^{(3)}$, $\mathcal{L}_{1,2}^{(3)}$, $\mathcal{L}_{1,3}^{(3)}$,
$\mathcal{L}_{2,1}^{(3)}$, $\mathcal{L}_{2,2}^{(3)}$, $\mathcal{L}_{2,3}^{(3)}$, $\mathcal{L}_{2,4}^{(3)}$,
$\mathcal{L}_{3,1}^{(3)}$, $\mathcal{L}_{3,2}^{(3)}$, $\mathcal{L}_{3,3}^{(3)}$, $\mathcal{L}_{3,4}^{(3)}$, $\mathcal{L}_{3,5}^{(3)}$,
$\mathcal{L}_{4,1}^{(3)}$, $\mathcal{L}_{4,2}^{(3)}$, $\mathcal{L}_{4,3}^{(3)}$, $\mathcal{L}_{4,4}^{(3)}$, $\mathcal{L}_{4,5}^{(3)}$,
$\mathcal{L}_{5,1}^{(3)}$, $\mathcal{L}_{5,2}^{(3)}$, $\mathcal{L}_{5,3}^{(3)}$, $\mathcal{L}_{5,4}^{(3)}$, $\mathcal{L}_{5,5}^{(3)}$, $\mathcal{L}_{5,6}^{(3)}$,
$\mathcal{L}_{6,1}^{(3)}$, $\mathcal{L}_{6,2}^{(3)}$, $\mathcal{L}_{6,3}^{(3)}$, $\mathcal{L}_{6,4}^{(3)}$, $\mathcal{L}_{6,5}^{(3)}$, $\mathcal{L}_{6,6}^{(3)}$ and $\mathcal{L}_{6,7}^{(3)}$,
respectively.
Namely $\mathcal{L}_{i,j}^{(3)}$ denotes the $j$-th subspace of the $i$-th equation.
The rank-3 Hilbert-Schmidt basis of $\mathcal{L}_{i,1}^{(3)}$, $i=1,$ 2, $\dots$, 6, can be constructed via any $4\times4$ isometric matrix.
That is, we can replace the entries of the $i$-th basis element by the entries in the $i$-th column of any $4$ by $4$ isometric matrix without zero entries, respectively.
Similarly, it is straightforward that $\mathcal{L}_{1,2}^{(3)}$,
 $\mathcal{L}_{2,2}^{(3)}$, $\mathcal{L}_{2,3}^{(3)}$,
 $\mathcal{L}_{3,2}^{(3)}$, $\mathcal{L}_{3,3}^{(3)}$, $\mathcal{L}_{3,4}^{(3)}$,
$\mathcal{L}_{4,2}^{(3)}$, $\mathcal{L}_{4,3}^{(3)}$, $\mathcal{L}_{4,4}^{(3)}$,
$\mathcal{L}_{5,2}^{(3)}$, $\mathcal{L}_{5,3}^{(3)}$, $\mathcal{L}_{5,4}^{(3)}$, $\mathcal{L}_{5,5}^{(3)}$,
$\mathcal{L}_{6,2}^{(3)}$, $\mathcal{L}_{6,3}^{(3)}$, $\mathcal{L}_{6,4}^{(3)}$, $\mathcal{L}_{6,5}^{(3)}$ and $\mathcal{L}_{6,6}^{(3)}$
have rank-3 Hilbert-Schmidt bases according to the existence of $3\times3$ isometric matrices.
It turns out that, for any given $i_0$, any rank-3 Hilbert-Schmidt basis of $\{\mathcal{L}_{i_0,j}^{(3)}: j=1,\dots, i_0+1 \ {\rm if}\ i_0\leq 3; j=1,\dots, i_0 \ {\rm if} \ i_0\geq 4\}$ form an unextendible rank-3 Hilbert-Schmidt basis of the associate matrix space, $i_0=1$, 2, $\dots$, 6.

Similarly, with the same spirit in mind, we can check that there are infinitely many UEBks (but not SUEBks) for any $k\geq4$ and any $d_1$ and $d_2$.
In general, $\mathcal{M}_{(k+r)\times (k+r')}$ can be decomposed as
\begin{eqnarray}
\mathcal{M}_{(k+r)\times (k+r')}=\mathcal{L}_1^{(k)}\oplus\mathcal{L}_2^{(k)}\oplus\cdots\oplus\mathcal{L}_{r+r'+1}^{(k)}
\oplus\mathcal{L}_{r+r'+2}^{(k)}\oplus\mathcal{L}_{r+r'+3}^{(k)},
\end{eqnarray}
where $\mathcal{L}_{l}^{(k)}$, $l=1$, 2, $\dots$, $r+r'+3$, satisfying the following:
\begin{itemize}
\item
For any $A=[a_{ij}]\in\mathcal{L}_{1}^{(k)}$, $a_{ij}=0$ whenever (1) $2\leq i\leq r+1$, or (2) $i=1$ and $j<k+r'$, or
(3) $i=r+1+s$ with $s<j$;
\item
If $2\leq l\leq r+2$, then for any $A=[a_{ij}]\in\mathcal{L}_{l}^{(k)}$, $a_{ij}=0$ whenever either $i\neq j+l-2$ and $j\leq k$ or $j>k$;
\item
If $r+3\leq l\leq r+r'+2$, then for any $A=[a_{ij}]\in\mathcal{L}_{l}^{(k)}$, $a_{ij}=0$ whenever either $j\neq i+l-r-2$ and $i\leq k$ or $i>k$.
\end{itemize}
The rank-k Hilbert-Schmidt basis of $\mathcal{L}_{1}^{(k)}$ can be constructed via any $k\times k$ isometric matrix.
Namely, we can replace the entries of the $i$-th basis element by the entries in the $i$-th column of any $k$ by $k$ isometric matrix without zero entries, respectively.
Similarly, $\mathcal{L}_{l}^{(k)}$, $2\leq l\leq r+r'+2$,
has rank-k Hilbert-Schmidt bases according to the existence of $k\times k$ isometric matrices.
Therefore, any rank-k Hilbert-Schmidt basis of $\{\mathcal{L}_{l}^{(k)}: 1\leq l\leq r+r'+2\}$ is an unextendible rank-k Hilbert-Schmidt basis of the associate matrix space since any Hilbert-Schmidt basis of $\mathcal{L}_{r+r'+3}^{(k)}$ is of rank smaller than $k$.
We thus obtain the following result.

\smallskip

\noindent{\bf Theorem 1}  {\it Infinite many UEBks  (but not SUEBks) exist in $d_1\otimes d_2$ systems for any $d_1$ and $d_2$, $k>1$.
}

\smallskip

It is worth mentioning here that all of the approaches of constructing UEBks above
lead to that any two matrices in the matrix space corresponding to the $n$-member UEBk
have $(d_1d_2-n)$ zero entries with the same entry position. We call it \emph{zero entries condition} for simplicity.
There also exist UEBks which do not admit the zero entries condition.
For example,
\begin{eqnarray}
A_1=\frac{2}{5}\left(\begin{array}{cc}
\frac{1}{2}&2\\
1&1\end{array}\right),
A_2=\frac{4}{\sqrt{73}}\left(\begin{array}{cc}
1&-\frac{5}{4}\\
1&1\end{array}\right),
A_3=\frac{21}{\sqrt{3650}}\left(\begin{array}{cc}
\frac{52}{21}&\frac{8}{21}\\
-1&-1\end{array}\right),
\end{eqnarray}
form an unextendible UEB2 for $2\otimes 2$ case. But the corresponding matrix space is
\begin{eqnarray}
\left(\begin{array}{cc}
*&*\\
a&a\end{array}\right), \forall \ a\in\mathbb{C},
\end{eqnarray}
which does not admit the zero entries condition.

\section{Constructing $(m+1)$-partite UEBk from $m$-partite UEBk}

We begin with the simple case of $k=1$, that is, the multipartite unextendible product basis.

\smallskip

\noindent{\bf Observation}  {\it If $\{|\psi_i\rangle\}$ is a UPB of $\mathbb{C}^{d_1}\otimes\mathbb{C}^{d_2}\otimes\cdots\otimes\mathbb{C}^{d_m}$,
then $\{|\psi_{i,j}\rangle\}$ is a UPB of $\mathbb{C}^{d_1}\otimes\mathbb{C}^{d_2}\otimes\cdots\otimes\mathbb{C}^{d_{m+1}}$,
where $|\psi_{i,j}\rangle=|\psi_i\rangle|j\rangle$, $j=0$, $1$, $\dots$, $d_m-1$.}

\smallskip

We denote by $|\psi^{l}\rangle:=|e_l^{(1)}\rangle|e_l^{(2)}\rangle\cdots|e_l^{(m)}\rangle$ for convenience provided that $|\psi\rangle=\sum_{j=0}^{k-1}\lambda_j|e_j^{(1)}\rangle|e_j^{(2)}\rangle\cdots|e_j^{(m)}\rangle$ with $\{|e_j^{(s)}\rangle\}$ is an orthonormal
set of $\mathbb{C}^{d_s}$, $s=1$, 2, $\dots$, $m$.
Let $\{|\psi_i\rangle\}$ be a $n$-member $m$-partite UEBk of $\mathbb{C}^{d_1}\otimes\mathbb{C}^{d_2}\otimes\cdots\otimes\mathbb{C}^{d_m}$, $1\leq k\leq d_1$.
We define
\begin{eqnarray}
|\psi_{i,j}^{(m+1)}\rangle:=\sum_{l=0}^{k-1}\lambda_l^{(i)}|\psi_i^{l}\rangle|j+l\rangle,
\label{mpartite}
\end{eqnarray}
where $\{|j\rangle\}$ is the standard computational basis of $\mathbb{C}^{d_{m+1}}$, $j=0$, 1, $\dots$, $d_{m+1}-1$,
$j+l$ means $j+l$ mod $d_{m+1}$, $i=0$, 1, $\dots$, $n$.
The following is the main result in this section.

\smallskip

\noindent{\bf Theorem 2}  {\it If $\{|\psi_i\rangle\}$ is an $m$-partite UEBk of $\mathbb{C}^{d_1}\otimes\mathbb{C}^{d_2}\otimes\cdots\otimes\mathbb{C}^{d_m}$, $1\leq k\leq d_1$,
then $\{|\psi_{i,j}^{(m+1)}\rangle\}$ defined as in Eq.~(\ref{mpartite}) is an $(m+1)$-partite UEBk of $\mathbb{C}^{d_1}\otimes\mathbb{C}^{d_2}\otimes\cdots\otimes\mathbb{C}^{d_{m+1}}$.
}

\smallskip

\noindent{\it Proof} \ Let $V_1$ be the subspace spanned by $\{|\psi_i\rangle\}$ and
let
$|\psi_{i}\rangle:=\sum_{l=0}^{k-1}\lambda_l^{(i)}|\psi_i^{l}\rangle$
be their Schmidt decomposition forms.
We denote the space spanned from $\{|\psi_{i,j}^{(m+1)}\rangle\}$ by $\check{V}_1$.
Then $\dim \check{V}_1=d_{m+1}\dim V_1$.
It follows that
$\dim \check{V}_1^\perp=d_{m+1}\dim V_1^\perp$.
Therefore $\check{V}_1=V_1\otimes\mathbb{C}^{d_{m+1}}$ and $\check{V}_1^\perp=V_1^\perp\otimes\mathbb{C}^{d_{m+1}}$.
This guarantees that any vector in $\check{V}_1^\perp$ either does not admit a Schmidt decomposition form or the
Schmidt number is not $k$.
In fact, for any $|\phi\rangle\in\check{V}_1^\perp$, if $\tilde{S}_r(|\phi\rangle)=k$,
we assume that $|\phi\rangle=\sum_{i=0}^{k-1}\lambda_i|\phi^i\rangle$.
Let $\rho_{\overline{m+1}}={\rm Tr}_{m+1}(|\phi\rangle\langle\phi|)$ and let the
spectral decomposition of $\rho_{\overline{m+1}}$ be
$\rho_{\overline{m+1}}=\sum_{i=0}^{k-1}\lambda_i^2|\phi_{(i)}\rangle\langle\phi_{(i)}|$.
(For $i\in\{1,2,3,\dots,n\}$, we denote by
$\bar{i}$ the combination consisting of all elements in
$\{1,2,\dots,n\}-\{i\}$, for instance, if $n=4$, $i=(2)$, then
$\bar{i}=(134)$.)
Then $|\phi_{(i)}\rangle\langle\phi_{(i)}|={\rm Tr}_{m+1}(|\phi^i\rangle\langle\phi^i|)$.
Let $|\phi'\rangle=\sum_{i=0}^{k-1}\lambda_i|\phi_{(i)}\rangle$, then $|\phi'\rangle\in V_1$,
which leads to $|\phi\rangle\in V_1\otimes \mathbb{C}^{m+1}$, a contradiction.
That is, $\{|\psi_{i,j}^{(m+1)}\rangle\}$ is unextendible.
The orthogonality and the Schmidt number condition are clear.
\hfill$\qed$

\smallskip

Theorem 2 implies that $(m+1)$-partite UEBk can be obtained from $m$-partite UEBk for any $m\geq2$.
Theorem 1 shows that there are infinitely many UEBks in any bipartite space, this guarantees the existence of UEBks in any multipartite
systems.

\smallskip

\noindent{\bf Theorem 3}  {\it There are infinitely many UEBks (but not SUEBks) in $\mathbb{C}^{d_1}\otimes\mathbb{C}^{d_2}\otimes\cdots\otimes\mathbb{C}^{d_m}$ with any dimensions and any $m\geq3$.
}

\smallskip

It is shown in Ref.~\cite{Bravyi} that there is no SUEB2 in $\mathbb{C}^{2}\otimes\mathbb{C}^{2}$. If $\{|\psi_i\rangle:i=1,2,\dots,n\}$ is a SUEB2 in $\mathbb{C}^{2}\otimes\mathbb{C}^{2}\otimes\mathbb{C}^{2}$,
then we can let $|\psi_i\rangle=U_i\otimes V_i\otimes W_i|\phi\rangle$ with
$|\phi\rangle=\frac{1}{\sqrt{2}}(|0\rangle|0\rangle|0\rangle+|1\rangle|1\rangle|1\rangle$,
where $U_i$, $V_i$ and $W_i$ are unitary, $\sum_{k=0}^1\langle k|\langle k|\langle k|(U_i^\dag U_j)\otimes(V_i^\dag V_j)\otimes (W_i^\dag W_j)|k\rangle|k\rangle|k\rangle=2\delta_{ij}$,
$\sum_{k=0}^1\langle k|\langle k|\langle k|(U_i^\dag U)\otimes(V_i^\dag V)\otimes (W_i^\dag W)|k\rangle|k\rangle|k\rangle=2\delta_{ij}$ implies $U$, $V$ and $W$ can not be unitary (or can not be unitary simultaneously), $i=1$, 2, $\dots$, $n$.
According to the proof of Lemma 1 in Ref.~\cite{Bravyi}, $2\times 2$ unitary matrices is always extendible, we conjecture that
there is no SUEB2 in $2^{\otimes d}$ systems, either.

\smallskip

Since there always exist SUEBks in any bipartite space $\mathbb{C}^{d_1}\otimes\mathbb{C}^{d_2}$ when $k<d_1\leq d_2$ and there exist UMEB when $d_1<d_2$ \cite{Bravyi,Chen,Limaosheng,Wangyanling,Guowu2014},
the following is clear.

\smallskip

\noindent{\bf Proposition 1}  {\it There are SUEBks in any multipartite system
$\mathbb{C}^{d_1}\otimes\mathbb{C}^{d_2}\otimes\cdots\otimes \mathbb{C}^{d_m}$ provided that $k< d_1$.
There exists UMEB in
${d}^{\otimes{m}}$ whenever there exists UMEB in $\mathbb{C}^{d}\otimes\mathbb{C}^{d}$.
}

\smallskip

There exists UMEB in $\mathbb{C}^{d}\otimes\mathbb{C}^{d}$ for $d=3,4,6$ \cite{Bravyi,Wangyanling}, so
there exists UMEB in
${d}^{\otimes{m}}$ for $d=3,4,6$.

We now discuss the structure of the subspace spanned by UEBk.
We consider the three-partite case $\mathbb{C}^{d_1}\otimes\mathbb{C}^{d_2}\otimes\mathbb{C}^{d_3}$
and the $m$-partite case with $m>3$ can be argued similarly.
Let $\{|\psi_i\rangle\}$ be a UEBk in $\mathbb{C}^{d_1}\otimes\mathbb{C}^{d_2}\otimes\mathbb{C}^{d_3}$
and $V_1$ be the subspace spanned by $\{|\psi_i\rangle\}$. Then $\mathbb{C}^{d_1}\otimes\mathbb{C}^{d_2}\otimes\mathbb{C}^{d_3}=V_1\oplus V_1^\bot$.
If $|\psi\rangle\in V_1$ admits a Schmidt decomposition form
$|\psi\rangle=\sum_{j=0}^{s}\lambda_j|x_j^{(1)}\rangle|x_j^{(2)}\rangle|x_j^{(3)}\rangle$, $s\leq k$,
then $|x_j^{(1)}\rangle|x_j^{(2)}\rangle|x_j^{(3)}\rangle\in V_1$ for any $1\leq j\leq s$.
We claim that $V_1^\bot$ does not contain vector with Schmidt number greater than $k$.
In order to see this, we
assume to reach a contradiction that $|\phi\rangle\in V_1^\bot$ and $|\phi\rangle=\sum_{j=0}^k\lambda_j|y_j^{(1)}\rangle|y_j^{(2)}\rangle|y_j^{(3)}\rangle$,
where $\{|y_j^{(i)}\rangle\}$ is an orthonormal set of $\mathbb{C}^{d_i}$, $i=1$, 2, 3.
Let
$|\phi'\rangle=\sum_{j=0}^{k-1}\lambda_j|y_j^{(1)}\rangle$
$|y_j^{(2)}\rangle|y_j^{(3)}\rangle$,
then $|\phi'\rangle\in V_1^\bot$. This implies that
$\{|\psi_i\rangle\}$ is extendible, a contradiction.
That is, $\mathbb{C}^{d_1}\otimes\mathbb{C}^{d_2}\otimes\mathbb{C}^{d_3}$ can be divided into two subspaces:
one contains vectors with Schmidt number which is equal to or greater than $k$ and the other one does not contain
vectors with Schmidt number which is equal to or greater than $k$.

\smallskip

\noindent{\bf Proposition 2}  {\it Let $\{|\psi_i\rangle\}$ be an $m$-partite UEBk in $\mathbb{C}^{d_1}\otimes\mathbb{C}^{d_2}\otimes\cdots \mathbb{C}^{d_m}$
and $V_1$ be the subspace spanned by $\{|\psi_i\rangle\}$.
Then $V_1^\bot$ does not contain vectors with Schmidt number is equal to or greater than $k$.
}

\section{Examples}

We only give several examples for three-partite UEBks, the $m$-partite UEBks ($1\leq k\leq d_1$) with $m\geq4$ can be deduced according to the method in Theorem 2.

\subsection{SUEB$k$}

Several examples of SUEBks for bipartite system are proposed in Refs.~\cite{Bennett1999,DiVincenzo,Bravyi,Chen,Limaosheng,Guowu2014}.
We hence can derive three-partite SUEBks by Theorem 2.

\subsubsection{k=1} We begin with the case of $k=1$, i.e., the $m$-partite UPB.

\smallskip

\noindent{\bf Example 1}  In $3\otimes 3$ space, two sets of UPB are proposed in Ref.~\cite{Bennett1999}.
We thus can obtain two sets of three-partite UPB in $3\otimes 3\otimes d$ straightforwardly for any $d\geq 2$.
We list them for the case of $d=2$.
\begin{eqnarray}
|\phi_{0,0}\rangle&=&\frac{1}{\sqrt{2}}|0\rangle(|0\rangle-|1\rangle)|0\rangle,\nonumber\\
|\phi_{0,1}\rangle&=&\frac{1}{\sqrt{2}}|0\rangle(|0\rangle-|1\rangle)|1\rangle,\nonumber\\
|\phi_{1,0}\rangle&=&\frac{1}{\sqrt{2}}|2\rangle(|1\rangle-|2\rangle)|0\rangle,\nonumber\\
|\phi_{1,1}\rangle&=&\frac{1}{\sqrt{2}}|2\rangle(|1\rangle-|2\rangle)|1\rangle,\nonumber\\
|\phi_{2,0}\rangle&=&\frac{1}{\sqrt{2}}(|0\rangle-|1\rangle)|2\rangle|0\rangle,\nonumber\\
|\phi_{2,1}\rangle&=&\frac{1}{\sqrt{2}}(|0\rangle-|1\rangle)|2\rangle|1\rangle,\nonumber\\
|\phi_{3,0}\rangle&=&\frac{1}{\sqrt{2}}(|1\rangle-|2\rangle)|0\rangle|0\rangle,\nonumber\\
|\phi_{3,1}\rangle&=&\frac{1}{\sqrt{2}}(|1\rangle-|2\rangle)|0\rangle|1\rangle,\nonumber\\
|\phi_{3,0}\rangle&=&\frac{1}{3}(|0\rangle+|1\rangle+|2\rangle)(|0\rangle+|1\rangle+|2\rangle)|0\rangle,\nonumber\\
|\phi_{3,1}\rangle&=&\frac{1}{3}(|0\rangle+|1\rangle+|2\rangle)(|0\rangle+|1\rangle+|2\rangle)|1\rangle
\end{eqnarray}
is a ten-member three-partite UPB from the TILES vectors in Ref.~\cite{Bennett1999}.
Let
\begin{eqnarray*}
\vec{v}_i=N(\cos \frac{2\pi i}{5}, \sin \frac{2\pi i}{5}, h ), \ i=0,1,\dots, 4
\end{eqnarray*}
with $h=\frac{1}{2}\sqrt{1+\sqrt{5}}$, $N=2/\sqrt{5+\sqrt{5}}$.
Then $|\psi_i\rangle=|\vec{v}_i\rangle|\vec{v}_{2i\ {\rm mod } \ 5}\rangle$, $i=0$, 1, $\dots$, 4,
form a UPB in $3\otimes 3$ space ~\cite{Bennett1999}.
Thus
\begin{eqnarray}
|\psi_{i,j}\rangle=|\psi_{i}\rangle|j\rangle,\ i=0, 1, \dots, 4, j=0,1
\end{eqnarray}
constitute a ten-member three-partite UPB in $3\otimes3\otimes2$ space.

\smallskip

\subsubsection{$1<k<d_1$}

According to Theorem 2, we can obtain $m$-partite UEBks from Propositions 1-6 and Eq.~(8) in \cite{Guowu2014}.
We list here one of them, the others can be followed straightforwardly.

\smallskip

\noindent{\bf Example 2}  Let
\begin{eqnarray}
|\phi_{mnls}\rangle:=\frac{1}{\sqrt{k}}\sum\limits_{p=0}^{k-1}\zeta_k^{np}|p\oplus m\rangle|(l-1)k+p\rangle|p+s\rangle,
\end{eqnarray}
where $m=0$, $1$, $\dots$, $d_1-1$, $n=0$, 1, $\dots$, $k-1$, $1<k<d_1$, $l=1$, $\dots$, $t$, $s=0$, 1, $\dots$, $d_3-1$,
$\zeta_k=e^\frac{2\pi\sqrt{-1}}{k}$ and $d_2=tk+r$, $0<r<k$, $x\oplus m$ denotes $x+m$ mod $d_1$, $p+s$ denotes $(p+s)$ mod $d_{3}$.
Then $\{|\phi_{mnls}\rangle\}$ is a $tkd_1d_3$-member UEBk in $\mathbb{C}^{d_1}\otimes\mathbb{C}^{d_2}\otimes\mathbb{C}^{d_3}$.

\smallskip

\subsubsection{$k=d_1$}

\noindent{\bf Example 3}  The following four pure states form an UMEB in $\mathbb{C}^{2}\otimes\mathbb{C}^{3}$ \cite{Chen}.
\begin{eqnarray*}
|\phi_0\rangle&=&\frac{1}{\sqrt{2}}(|0\rangle|0\rangle+|1\rangle|1\rangle),\nonumber\\
|\psi_i\rangle&=&(\sigma_i\otimes I_3)|\phi_0\rangle,\ i=1,2,3.
\end{eqnarray*}
Let $|\psi_i^j\rangle=(\sigma_i\otimes I_3)|j\rangle|j\rangle$, $j=0$, 1, and
\begin{eqnarray}
|\phi_{0,0}\rangle&=&\frac{1}{\sqrt{2}}(|0\rangle|0\rangle|0\rangle+|1\rangle|1\rangle|1\rangle),\nonumber\\
|\phi_{0,1}\rangle&=&\frac{1}{\sqrt{2}}(|0\rangle|0\rangle|1\rangle+|1\rangle|1\rangle|2\rangle),\nonumber\\
|\phi_{0,2}\rangle&=&\frac{1}{\sqrt{2}}(|0\rangle|0\rangle|2\rangle+|1\rangle|1\rangle|0\rangle),\nonumber\\
|\psi_{i,0}\rangle&=&\frac{1}{\sqrt{2}}(|\psi_i^0\rangle|0\rangle+|\psi_i^1\rangle|1\rangle),\nonumber\\
|\psi_{i,1}\rangle&=&\frac{1}{\sqrt{2}}(|\psi_i^0\rangle|1\rangle+|\psi_i^1\rangle|2\rangle),\nonumber\\
|\psi_{i,2}\rangle&=&\frac{1}{\sqrt{2}}(|\psi_i^0\rangle|2\rangle+|\psi_i^1\rangle|0\rangle),
 i=1,2,3.
\end{eqnarray}
Then $\{|\phi_{0,j}\rangle, |\psi_{i,j}\rangle: i=1,2,3, j=0,1,2\}$ is a $12$-member 3-partite
UMEB in $\mathbb{C}^{2}\otimes\mathbb{C}^{3}\otimes\mathbb{C}^{3}$.

From Propositions 1-2 in Ref.~\cite{Limaosheng}, one can obtain two types of
three-partite UMEBs for the case of $d_1<d_2$.

\subsection{UEB$k$ but not SUEB$k$}

We present here examples of three-partite UEBks by Theorem 2 from the method in Section 3 for the bipartite UEBks.

\smallskip

\noindent{\bf Example 4}  We rewrite the vectors in Eq.~(\ref{2qubits}) in the Schmidt decomposition form as
\if is
\begin{eqnarray*}
\left\{\frac{1}{3}\left(\begin{array}{cc}
-1&0\\
2&2\end{array}\right)\right\},
\left\{\frac{1}{3}\left(\begin{array}{cc}
2&0\\
-1&2\end{array}\right)\right\},
\left\{\frac{1}{3}\left(\begin{array}{cc}
2&0\\
2&-1\end{array}\right)\right\}.
\end{eqnarray*}
The singular value decompositions of the matrices above are
, the following six vectors form a UEBk in three-qubits system.
\fi
\begin{eqnarray}
|\psi_0\rangle&=&\sqrt{\frac{746}{787}}(\frac{664}{2587}|0\rangle+\frac{577}{597}|1\rangle)(\frac{1165}{1554}|1\rangle-\frac{771}{1165}|0\rangle),\nonumber\\
              &&-\sqrt{\frac{282}{5413}}(\frac{664}{2587}|0\rangle-\frac{577}{597}|0\rangle)(\frac{1165}{1554}|0\rangle+\frac{771}{1165}|0\rangle),\nonumber\\
|\psi_1\rangle&=&\sqrt{\frac{1445}{1982}}(\frac{1243}{1577}|1\rangle-\frac{1717}{2790}|0\rangle)
(\frac{1717}{2790}|1\rangle-\frac{1243}{1577}|0\rangle),\nonumber\\
              &&-\sqrt{\frac{537}{1982}}(\frac{1243}{1577}|0\rangle+\frac{1717}{2790}|1\rangle)
              (\frac{1717}{2790}|0\rangle+\frac{1243}{1577}|1\rangle),\nonumber\\
|\psi_2\rangle&=&-\sqrt{\frac{746}{787}}(\frac{771}{1165}|0\rangle+\frac{1165}{1554}|1\rangle)
                                          (\frac{664}{2587}|0\rangle+\frac{577}{597}|1\rangle),\nonumber\\
              &&-\sqrt{\frac{282}{5413}}(\frac{771}{1165}|1\rangle-\frac{1165}{1554}|0\rangle)
                                         (\frac{664}{2587}|0\rangle-\frac{577}{597}|0\rangle).
\label{3-2qubits}
\end{eqnarray}
It turns out that
\begin{eqnarray}
|\psi_{0,0}\rangle&=&\sqrt{\frac{746}{787}}(\frac{664}{2587}|0\rangle+\frac{577}{597}|1\rangle)
              (\frac{1165}{1554}|1\rangle-\frac{771}{1165}|0\rangle)|0\rangle,\nonumber\\
              &&-\sqrt{\frac{282}{5413}}(\frac{664}{2587}|0\rangle-\frac{577}{597}|0\rangle)
              (\frac{1165}{1554}|0\rangle+\frac{771}{1165}|0\rangle)|1\rangle,\nonumber\\
|\psi_{0,1}\rangle&=&\sqrt{\frac{746}{787}}(\frac{664}{2587}|0\rangle+\frac{577}{597}|1\rangle)
              (\frac{1165}{1554}|1\rangle-\frac{771}{1165}|0\rangle)|1\rangle,\nonumber\\
              &&-\sqrt{\frac{282}{5413}}(\frac{664}{2587}|0\rangle-\frac{577}{597}|0\rangle)
              (\frac{1165}{1554}|0\rangle+\frac{771}{1165}|0\rangle)|0\rangle,\nonumber\\
|\psi_{1,0}\rangle&=&\sqrt{\frac{1445}{1982}}(\frac{1243}{1577}|1\rangle-\frac{1717}{2790}|0\rangle)
              (\frac{1717}{2790}|1\rangle-\frac{1243}{1577}|0\rangle)|0\rangle,\nonumber\\
              &&+\sqrt{\frac{537}{1982}}(\frac{1243}{1577}|0\rangle+\frac{1717}{2790}|1\rangle)
              (\frac{1717}{2790}|0\rangle+\frac{1243}{1577}|1\rangle)|1\rangle,\nonumber\\
|\psi_{1,0}\rangle&=&\sqrt{\frac{1445}{1982}}(\frac{1243}{1577}|1\rangle-\frac{1717}{2790}|0\rangle)
              (\frac{1717}{2790}|1\rangle-\frac{1243}{1577}|0\rangle)|1\rangle,\nonumber\\
              &&+\sqrt{\frac{537}{1982}}(\frac{1243}{1577}|0\rangle+\frac{1717}{2790}|1\rangle)
              (\frac{1717}{2790}|0\rangle+\frac{1243}{1577}|1\rangle)|0\rangle,\nonumber\\
|\psi_{2,0}\rangle&=&-\sqrt{\frac{746}{787}}(\frac{771}{1165}|0\rangle+\frac{1165}{1554}|1\rangle)
                                          (\frac{664}{2587}|0\rangle+\frac{577}{597}|1\rangle)|0\rangle,\nonumber\\
              &&+\sqrt{\frac{282}{5413}}(\frac{771}{1165}|1\rangle-\frac{1165}{1554}|0\rangle)
                                         (\frac{664}{2587}|0\rangle-\frac{577}{597}|0\rangle)|1\rangle,\nonumber\\
|\psi_{2,1}\rangle&=&-\sqrt{\frac{746}{787}}(\frac{771}{1165}|0\rangle+\frac{1165}{1554}|1\rangle)
                                          (\frac{664}{2587}|0\rangle+\frac{577}{597}|1\rangle)|1\rangle,\nonumber\\
              &&+\sqrt{\frac{282}{5413}}(\frac{771}{1165}|1\rangle-\frac{1165}{1554}|0\rangle)
                                         (\frac{664}{2587}|0\rangle-\frac{577}{597}|0\rangle)|0\rangle
\label{3qubits}
\end{eqnarray}
form a three-partite UEB2 in $\mathbb{C}^2\otimes\mathbb{C}^2\otimes\mathbb{C}^2$.

\smallskip

\noindent{\bf Example 5}
It is straightforward that
\begin{eqnarray}
|\psi_0\rangle&=&\frac{1}{2}|0\rangle|0\rangle+\frac{\sqrt{3}}{2}|1\rangle|1\rangle,\nonumber \\
|\psi_1\rangle&=&\frac{\sqrt{3}}{2}|0\rangle|0\rangle-\frac{1}{2}|1\rangle|1\rangle,\nonumber \\
|\psi_2\rangle&=&\frac{1}{2}|0\rangle|1\rangle+\frac{\sqrt{3}}{2}|1\rangle|0\rangle,\nonumber \\
|\psi_3\rangle&=&\frac{\sqrt{3}}{2}|0\rangle|1\rangle-\frac{1}{2}|1\rangle|0\rangle
\label{ueb23}
\end{eqnarray}
constitute a UEB2 in $\mathbb{C}^2\otimes\mathbb{C}^3$, but it is not a SUEB2.
Hence, the following 12 vectors constitute a UEB2 but not SUEB2 in $\mathbb{C}^2\otimes\mathbb{C}^3\otimes\mathbb{C}^3$:
\begin{eqnarray}
|\psi_{0,0}\rangle&=&\frac{1}{2}|0\rangle|0\rangle|0\rangle+\frac{\sqrt{3}}{2}|1\rangle|1\rangle|1\rangle,\nonumber \\
|\psi_{0,1}\rangle&=&\frac{1}{2}|0\rangle|0\rangle|1\rangle+\frac{\sqrt{3}}{2}|1\rangle|1\rangle|2\rangle,\nonumber \\
|\psi_{0,2}\rangle&=&\frac{1}{2}|0\rangle|0\rangle|2\rangle+\frac{\sqrt{3}}{2}|1\rangle|1\rangle|0\rangle,\nonumber \\
|\psi_{1,0}\rangle&=&\frac{\sqrt{3}}{2}|0\rangle|0\rangle|0\rangle-\frac{1}{2}|1\rangle|1\rangle|1\rangle,\nonumber \\
|\psi_{1,1}\rangle&=&\frac{\sqrt{3}}{2}|0\rangle|0\rangle|1\rangle-\frac{1}{2}|1\rangle|1\rangle|2\rangle,\nonumber \\
|\psi_{1,2}\rangle&=&\frac{\sqrt{3}}{2}|0\rangle|0\rangle|2\rangle-\frac{1}{2}|1\rangle|1\rangle|0\rangle,\nonumber \\
|\psi_{3,0}\rangle&=&\frac{1}{2}|0\rangle|1\rangle|0\rangle+\frac{\sqrt{3}}{2}|1\rangle|0\rangle|1\rangle,\nonumber \\
|\psi_{3,0}\rangle&=&\frac{1}{2}|0\rangle|1\rangle|1\rangle+\frac{\sqrt{3}}{2}|1\rangle|0\rangle|2\rangle,\nonumber \\
|\psi_{3,0}\rangle&=&\frac{1}{2}|0\rangle|2\rangle|0\rangle+\frac{\sqrt{3}}{2}|1\rangle|0\rangle|0\rangle,\nonumber \\
|\psi_{4,0}\rangle&=&\frac{\sqrt{3}}{2}|0\rangle|1\rangle|0\rangle-\frac{1}{2}|1\rangle|0\rangle|1\rangle,\nonumber \\
|\psi_{4,1}\rangle&=&\frac{\sqrt{3}}{2}|0\rangle|1\rangle|1\rangle-\frac{1}{2}|1\rangle|0\rangle|2\rangle,\nonumber \\
|\psi_{4,2}\rangle&=&\frac{\sqrt{3}}{2}|0\rangle|1\rangle|2\rangle-\frac{1}{2}|1\rangle|0\rangle|0\rangle.
\label{ueb233}
\end{eqnarray}

There are many ways of constructing UEBks in multipartite case. For example,
the $2\otimes 3\otimes 3$ case, we can obtain a three-partite UEB2 either from a UEB2 in $2\otimes 3$
or from a UEB2 in $3\otimes 3$.

\section{Conclusion and discussion}

The multipartite UEBk was put forward, which extended the concepts of bipartite UEBk.
We proposed a general method of constructing $(m+1)$-partite UEBk from the
$m$-partite UEBk and showed that multipartite UEBk exists in any multipartite systems with any dimensions.
We listed several examples of multipartite SUEBks based on the bipartite examples proposed in the early papers.
By now, the unextendible basis problem is settled down.
Going further, the MUBs constitute now a basic ingredient in many applications
of quantum information processing.
The bipartite UMEBs can extend to two mutually unbiased basis (MUB) \cite{Chen,Fei2015}.
We thus can obtain multipartite mutually unbiased basis from multipartite UEBks.
The bipartite MUB has shown be useful in
investigating quantum state
tomography and cryptographic protocols.
We hope that the multipartite UEBks would also be useful in
studying quantum state
tomography and cryptographic protocols associated with multipartite quantum systems.

Unextendible entangled basis is not only a physical concept but also of great importance in
mathematics.
It reveals the structure of the tensor product of the Hilbert spaces.
For example, for any fixed Schmidt number $k$,
any tensor product of the Hilbert spaces can be divided into two subspaces:
one contains vectors with Schmidt number $k$ while the other only contains vectors with
Schmidt number smaller than $k$.

\begin{acknowledgements}
The authors wish to give their thanks to the referees for their helpful comments and
suggestions to improve the manuscript.
This work is supported by the National Natural Science Foundation of China under Grants No. 11301312 and 11171249,
the Natural Science Foundation of Shanxi
under Grant No. 2013021001-1 and the Research start-up fund for Doctors of Shanxi Datong University
under Grant No. 2011-B-01.
\end{acknowledgements}

\bibliographystyle{spmpsci}      


\end{document}